\documentclass[11pt]{article}
\usepackage{psfig}
\usepackage{citesort}

\newtheorem{theorem}{Theorem}
\newtheorem{lemma}[theorem]{Lemma}
\newtheorem{corollary}[theorem]{Corollary}
\newtheorem{definition}[theorem]{Definition}
\newtheorem{myrule}{Rule}

\newenvironment{proof}{\noindent\par{\bf Proof: }}{\nopagebreak\rule{1
ex}{0.8 em}\medskip}

\newcommand{\Prm}{\Pr\nolimits_m}
\newcommand{\overlineX}[0]{Y}
\newcommand{\overlineA}[0]{D}
\newcommand{\limm}[0]{\lim_{m \goesto \infty}}
\newcommand{\newloglike}[2]{\newcommand{#1}{\mathop{\rm #2}\nolimits}}
\newloglike{\sgn}{sign}

\newcommand{\trans}{\top}
\newcommand{\buzz}[1]{\emph{#1}}
\newcommand{\sss}{\{-1,0,+1\}}

\newcommand{\MM}{{\cal M}}
\newcommand{\FF}{{\cal F}}
\newcommand{\CC}{{\cal C}}
\newcommand{\SSS}{{\cal S}}

\newcommand{\newclass}[2]{\newcommand{#1}{\mbox{\rm #2}}}
\newclass{\pp}{PP}
\newclass{\np}{NP}
\newclass{\p}{P}
\newclass{\bcpp}{BCPP}
\newclass{\cpp}{CPP}
\newclass{\zpp}{ZPP}
\newclass{\bpp}{BPP}
\newclass{\crclass}{CR}
\newcommand{\pnplog}{\mbox{$\p{}^{\mbox{\scriptsize NP}[O(\log n)]}$}}

\newcommand{\goesto}{\rightarrow}

\newcommand{\suppress}[1]{}

\begin{document}

\title{Towards Understanding the Predictability of Stock Markets from
the Perspective of Computational Complexity\thanks{A preliminary
version will appear in {\em Proceedings of the 12th Annual ACM-SIAM
Symposium on Discrete Algorithms}, 2001.}}

\author{
James Aspnes\thanks{
Department of Computer Science, Yale University,
New Haven, CT 06520-8285, USA.
Email: {\tt aspnes@cs.yale.edu}.
Supported in part by NSF Grant CCR-9820888.}
\and 
David F. Fischer\thanks{Class of 1999, Yale College, New Haven, CT
06520-8285, USA. Email: {\tt fischer@aya.yale.edu}.}
\and
Michael J. Fischer\thanks{Department of Computer Science, Yale University,
New Haven, CT 06520-8285, USA.  Email: {\tt fischer-michael@cs.yale.edu}.} 
\and
Ming-Yang Kao\thanks{Department of Computer Science, Yale University,
New Haven, CT 06520-8285, USA.  Email: {\tt kao-ming-yang@cs.yale.edu}. 
Supported in part by NSF Grants CCR-9531028 and CCR-9988376.
Current address: Department of Electrical Engineering and Computer Science,
Tufts University, Medford, MA 02155, USA. Email: {\tt kao@eecs.tufts.edu}.}
\and
Alok Kumar\thanks{
School of Management,
Yale University,
New Haven, CT 06520, USA.
Email: {\tt alok.kumar@yale.edu}.}
}

\maketitle

\begin{abstract}
This paper initiates a study into the century-old issue of market
predictability from the perspective of computational complexity.  We
develop a simple agent-based model for a stock market where the agents
are traders equipped with simple trading strategies, and their trades
together determine the stock prices.  Computer simulations show that a
basic case of this model is already capable of generating price graphs
which are visually similar to the recent price movements of high tech
stocks. In the general model, we prove that if there are a large
number of traders but they employ a relatively small number of
strategies, then there is a polynomial-time algorithm for predicting
future price movements with high accuracy. On the other hand, if the
number of strategies is large, market prediction becomes complete in
two new computational complexity classes {\cpp} and {\bcpp}, where
$\pnplog\subseteq\bcpp\subseteq\cpp=\pp$.  These computational
completeness results open up a novel possibility that the price graph
of an actual stock could be sufficiently deterministic for various
prediction goals but appear random to all polynomial-time prediction
algorithms.
\end{abstract}


\section{Introduction} 
The issue of market predictability has been debated for more than a
century (see \cite{Cootner64} for earlier papers and
\cite{malkiel90,Fama91,Campbell97,Lo99} for more recent viewpoints).  
In 1900, the pioneering work ``{Theory of Speculation}'' of Louis
Bachelier used Brownian motion to analyze the stochastic properties of
security prices \cite{Cootner64}. Since then, Brownian motion and its
variants have become textbook tools for modeling financial assets.
Relatively recently, the radically different methodology of Mandelbrot
used fractals to approximate price graphs deterministically
\cite{mandelbrot97}.  In this paper, we initiate a study into this
long-running issue from the perspective of computational complexity.

We develop a simple agent-based model for a stock market
{\cite{Day90,Lux98}}.  The agents are traders equipped with
simple trading strategies, and their trades together determine the
stock prices. We first consider a basic case of this model where there
are only two strategies, namely, momentum and contrarian strategies.
The choice of this base model and thus our general model is justified
at two levels: (1) Experimental and empirical studies in the finance
literature
\cite{Andreassen90,Clarke98,DeBondt91,DeBondt93,Kumar99,Clarke98,Black86,DeLong90b} 
show that a large number of traders primarily follow these two
strategies.  (2) Our own simulation results show that despite its
simplicity, the base model is capable of generating price graphs which
are visually similar to the recent price movements of high tech stocks
(Figures \ref{figEx1} and \ref{figEx2}).

With these justifications, we then consider the issue of market
predictability in the general model.  We prove that if there are a
large number of traders but they employ a relatively small number of
strategies, then there is a polynomial-time algorithm to predict
future price movements with high accuracy (Theorem
\ref{theorem-prediction-easy-fixed}).  On the other hand, if there are
also a large number of strategies, then the problem of predicting
future prices becomes computationally very hard.  To describe this
hardness, we define two new computational complexity classes called
{\cpp} and {\bcpp} (Definitions \ref{def-cpp} and
\ref{def-bcpp}). We show that some market prediction problems are
complete for these two classes (Theorems \ref{theorem-bcpp-complete}
and \ref{theorem-cpp-complete}) and that
$\pnplog\subseteq\bcpp\subseteq\cpp=\pp$.

These computational completeness results open up the possibility that
the price graph of an actual stock could be sufficiently deterministic
for various prediction purposes but appear random to all
polynomial-time prediction algorithms. This is in contrast to the most
popular academic belief that the future price of a stock cannot be
predicted from its historical prices because the latter are
statistically random and contain no information. This new viewpoint
also differs from the fractal-based methodology in that the price
graph of a stock could be a fractal but the fractal might not be
computable in polynomial time.  The findings in this paper can by no
means settle the debate on market predictability. Our goal is only
that this alternative approach could provide new insights to the
predictability issue in a systematic manner.  In particular, it could
provide a general framework to investigate the many documented
technical trading rules \cite{GSW99} and to generate novel and
significant interdisciplinary research problems for computer science
and finance.

The rest of the paper is organized as follows.
Section~\ref{sec_experimental} discusses the basic market model.
Section~\ref{section-market-model} formulates the general model.
Section~\ref{section-prediction} proves the complexity results for
market prediction in the general model. We conclude the paper with
some directions for future research in Section~\ref{sec_conclusion}.

\section{A Basic Market Model}\label{sec_experimental}
In this section, we present a very simple market model, called the
{\it deterministic-switching MC} (DSMC) model. The letter M stands for
a {\it momentum} strategy, and the letter C for a {\it
contrarian} strategy.  These two strategies and the model
itself are defined in Section~\ref{subsec_basic_def}. Some computer
simulations for this model are reported in Section~\ref{subsec_exp}.

Intuitively, these strategies are heuristics (``rules of thumb'')
used by traders in the absence of reliable asset valuation models.  As
discussed in \cite{DeLong90b}, a momentum trader may observe a
sequence of ``up'' trades (price increments) and execute a buy
trade in the anticipation that she will not be one of the last buyers,
knowing very well that the asset is overpriced. Similarly, she may see
some ``down'' trades (price decrements) and then make a sell trade
in the hope that there will be more sellers after her.  In contrast,
after detecting a number of ``up'' (respectively, down) trades, a
contrarian trader may submit a sell (respectively, buy) trade,
anticipating a price reversal.

Both experimental and empirical studies have shown that traders look
at past price dynamics to form their expectations of future prices,
and a large number of them primarily follow momentum or contrarian
strategies \cite{Andreassen90,Clarke98,DeBondt91,DeBondt93}.  In
addition, the traders may switch between these two diametrically
opposite strategies.  Momentum and contrarian strategies are dominant
in the behavior of professional market timers as well
\cite{Kumar99}.  The use of momentum and contrarian strategies
sometimes signifies gambling tendencies among traders
\cite{Clarke98}. In fact, a market model with momentum and contrarian
traders can also be interpreted as a market with noise traders and
rational traders, where the noise traders essentially follow a
momentum strategy while the rational traders attempt to exploit the
noise traders by following a contrarian strategy
\cite{Black86,DeLong90b}.

\subsection{Defining the DSMC Model}\label{subsec_basic_def}
In the DSMC model, there is only one stock traded in the market.  The
model is completely specified by three integer parameters $m, L, k >
0$, and a real parameter $\alpha > 0$ as follows.

There are $m$ traders in the market, and each trader's strategy set
consists of momentum $(\MM)$ and contrarian $(\CC)$ strategies.  At
the beginning of day 1 of the investment period, each trader randomly
chooses her initial strategy from $\{\MM,\CC\}$ and an integer $\ell_i
\in [2,L]$ with equal probability, where $L$ is the {\it maximum
strategy switching period}.  This is the only source of randomness in
the DSMC model; from this point onwards, there is no random choice.

\begin{myrule}{\bf(Deterministic Strategy Switching Rule)} \rm    
For days $1,\ldots,k+1$, there is no trading.  Each trader starts
trading from day $k+2$ using her initial strategy.  Trader $i$ uses the
same strategy for $\ell_i$ days and switches it at the beginning of
every $\ell_i$ days.
\end{myrule}

The next rule defines the two strategies with respect to a given
memory size $k$, which is the same for all traders.

\newcommand{\TR}[0]{{\rm Tr}}

\begin{myrule}{\bf(Trading Rule)}\rm  
\label{rule_trading}
At the beginning of day $t$, observe the stock prices $P_f$ of days $f
\in [t-(k+1),t-1]$.  For $g \in [t-k,t-1]$, count the number $k_u$ of
days $g$ when $P_{g}>P_{g-1}$; and the number $k_d$ of days when
$P_{g} < P_{g-1}$.  The {\it $k$-day trend} is defied as $\TR(k,t) =
k_u - k_d$.  Then, if $\TR(k,t) \geq 0$ $($respectively, $<0)$, the
momentum strategy $\MM$ buys (respectively, sells) one share of the
stock at the market price determined by Rule~\ref{rule_price}
below. In contrast, the contrarian strategy $\CC$ sells (respectively,
buys) one share of the stock.
\end{myrule}

For instance, suppose that $k = 2$, and investor $i$ picks her initial
strategy $\MM$ and $\ell_i = 2$ at the beginning of day 1.  She then
observes the prices of days 1, 2, 3, which are, say, $\$80, \$82,
\$90$. At the beginning of day 4, she  issues a market order to buy
one share of the stock. The orders issued by the traders on day 4
together determine the price of day 4 as specified by Rule
\ref{rule_price}. Suppose that the price of day 4 is $\$91$,
then investor $i$ issues another market buy order at the beginning of
day 5. Since her $\ell_i$ is 2, at the beginning of day 6, she 
switches her strategy from $\MM$ to $\CC$.

\begin{myrule}{\bf(Price Adjustment Rule)}\rm  
\label{rule_price}
The prices for days $1,\ldots,k+1$ are given.  On day $t \geq k+2$,
let $m_b$ and $m_s$ be the total numbers of buys and sells,
respectively.  Then, the price $P_t$ on day $t$ is determined by the
following equation:
\[P_t - P_{t-1} = \alpha{\cdot}(m_b - m_s),\]
where $\alpha$ is the unit of price change.
\end{myrule}

\newlength{\heightone}
\setlength{\heightone}{0in}
\addtolength{\heightone}{0.5\textwidth}
\begin{figure}
\centerline{\psfig{figure=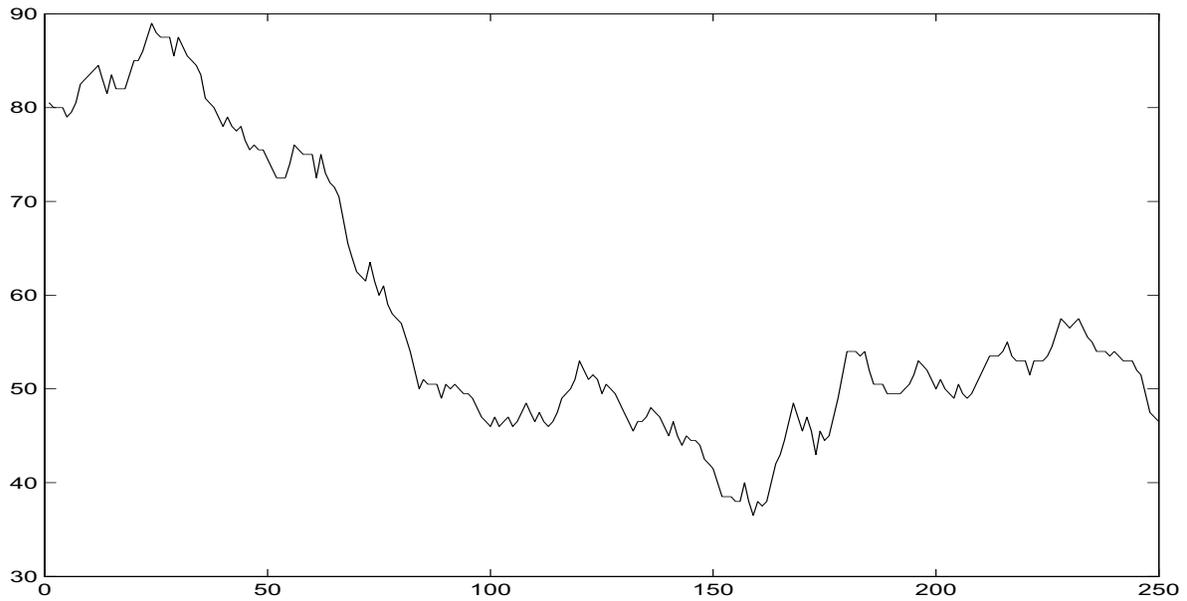,width=\textwidth,height=\heightone}}
\caption{A one-year price sequence generated using the DSMC model.
Parameters: number of traders $m = 20$, memory size $k=2$, maximum
strategy switching period $L=8$, unit of price change $\alpha = 0.25$,
number of trading days $= 250$.  The price graph appears strikingly
similar to the recent price movements of high tech stocks.}
\label{figEx1}
\end{figure}

\begin{figure}
\centerline{\psfig{figure=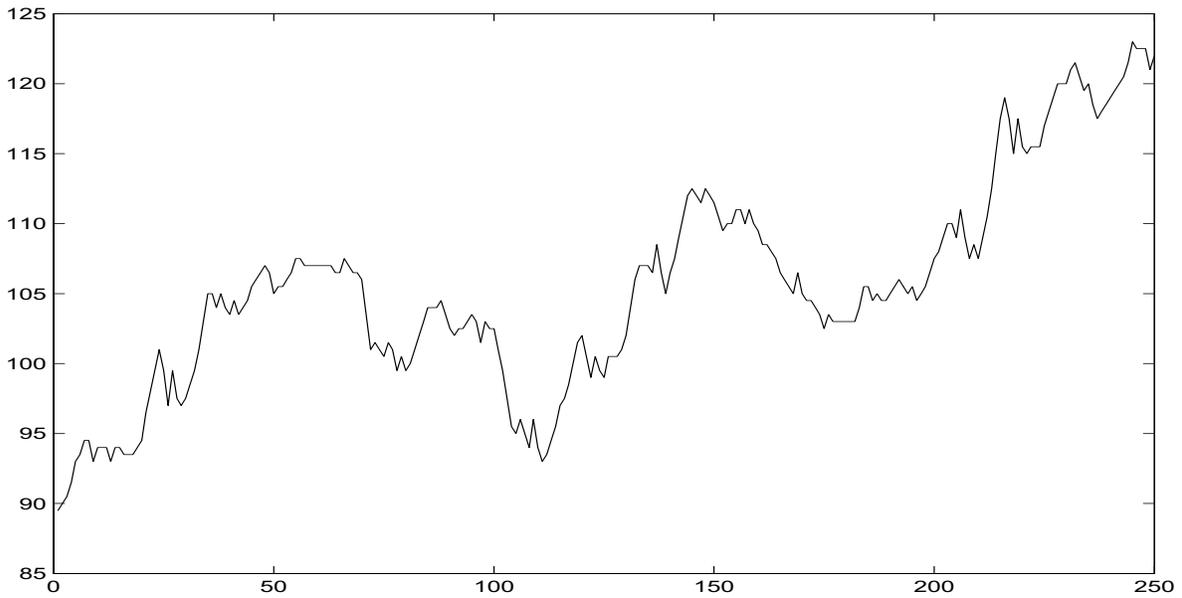,width=\textwidth,height=\heightone}}
\caption{A one-year price sequence generated using the DSMC model.
The parameters are the same as those for
Figure~\protect{\ref{figEx1}}.}
\label{figEx2}
\end{figure}

\subsection{Computer Simulation on the DSMC Model}\label{subsec_exp}
\label{subsec_simulate}
We have conducted some computer simulations of the DSMC model to test
whether it can generate realistic price graphs.  Because we had to
examine the graphs visually, our time constraints limited the number
of these simulations to only about six hundred.  For a large fraction
of them, we set $m = 20$, $L = 8$, and the initial $k$ prices in the
range of $\$70$ to $\$90$. We then focused on testing the effect of
memory size $k$ \cite{Mullainathan98}. Two main findings are as
follows:
\begin{itemize}
\item 
For $k=1$, the price graphs were not visually real.
\item 
For $k=2$, about one out of four graphs were strikingly similar to
those of recent high tech stocks, which was a major positive surprise
to us.  Two representatives of such graphs are shown in
Figures~\ref{figEx1} and \ref{figEx2}.
\end{itemize}

These two statements are based on our subjective impressions and
limited simulations. To further understand the DSMC model, it would be
useful to automate statistical analysis on the price graphs generated
by this model and compare them with real stock prices.

\section{A General Market Model}\label{section-market-model}
In this section, we define a market model, called the {\it AS} model,
where the word AS stands for arbitrary strategies.  It can be verified
in a straightforward manner that the DSMC model is a special case of
the AS model.

In the AS model, there is only one stock traded in the market.  The
model is completely specified as follows with five parameters: (1) the
number $m$ of traders, (2) a unit $\alpha > 0$ of price change, (3) a
set $\Pi=\{\SSS^1,\ldots,\SSS^h\}$ of strategies, (4) a price
adjustment rule (Equation~\ref{eq-proportional-increment} or
\ref{eq-fixed-increment} below), and (5) a joint distribution of the
population variables $X_1,\ldots,X_h$.

\begin{myrule}{\bf(Market Initialization)} \rm 
There are $m$ traders in the market.  At the beginning of day 1 of the
investment period, each trader randomly chooses her initial strategy
from $\Pi$.  Let $X_i$ be the number of traders who choose $\SSS^i$.
Then, each $X_i$ is a random variable, which is the only source of
randomness in the model. (Unlike the DSMC model, because the allowable
generality of $\Pi$, the AS model does not need strategy switching.)
\end{myrule} 

Different joint distributions of the variables $X_i$ lead to different
specific models and prediction problems.  In
Section~\ref{section-prediction-easy}, we consider joint distributions
that tend to Gaussian in the limit as the number $m$ of traders
becomes large. In Section~\ref{section-prediction-hard}, we consider
the case where the variables $X_i$ are independent, and each is $0$ or
$1$ with equal probability.

\begin{myrule}{\bf(Trading Strategies)}\rm\label{section-traders} 
There is no trading on day $0$. At the beginning of day $t \geq 1$, a
trader observes the historical prices $P_0,\ldots,P_{t-1}$ and reacts
by issuing a market order to buy one share of the stock, hold (i.e.,
do nothing), or sell one share according her strategy.  Formally, a
{\it strategy} is a collection of functions
$\SSS=\{\SSS_1,\SSS_2,\ldots,\SSS_t,\ldots\}$, where each $\SSS_t$
maps $P_0, \ldots, P_{t-1}$ to $+1$ (buy), $0$ (hold), or $-1$ (sell).
\end{myrule}

The price $P_t$ of day $t$ is determined at the end of the day by the
day's $m$ market orders using Rule~\ref{rule_price_as}. Since the
traders choose their strategies randomly, the sequence
$P_0,P_1,\ldots,P_t,\ldots$ is a stochastic process.  We write $\FF_t$
for the probability space induced by all possible sequences $\langle
P_0,\ldots,P_t\rangle$ \cite{KS98}.  Then, we think of each function
$\SSS_t$ as a random variable on $\FF_{t-1}$.

We distinguish between strategies that react to price movements and
those that ignore them.
\begin{itemize}
\item
$\SSS$ is an {\it active} strategy if the functions $\SSS_t$ may or
may not be constant functions.  An {\it active} trader is one with an
active strategy.
\item 
$\SSS$ is a {\it passive} strategy if the functions $\SSS_t$ all are
constant functions. A {\it passive} trader is one with a passive
strategy.
\end{itemize}

\begin{myrule}\rm {\bf (Price Adjustment)}\label{rule_price_as}
The price $P_0$ is given.  At the end of day $t \geq 1$, the price
$P_t$ is determined by the day's market orders to buy or sell from the
traders.  We consider two simple rules:

With the \buzz{proportional increment} (PI) rule,
\begin{equation}
\label{eq-proportional-increment}
P_t = P_{t-1} + \alpha{\cdot}\sum_{i=1}^{h} X_i{\cdot}\SSS^i_t,
\end{equation}
where $\alpha$ is the unit of price change.  Thus we can observe
directly the net difference between the number of buyers and sellers
on day $t$.

With the \buzz{fixed increment} (FI) rule,
\begin{equation}
\label{eq-fixed-increment}
P_t = P_{t-1} + \alpha{\cdot}\sgn\left(\sum_{i=1}^{h}
X_i{\cdot}\SSS^i_t\right).
\end{equation}
In this case, the market moves up or down depending on whether the
majority of traders are buying or selling, but the amount by which it
moves is fixed at $\alpha$.
\end{myrule}

For notational brevity, an {\it AS+FI model} refers to an AS model with the
fixed increment rule, and an {\it AS+PI model} refers to an AS model with
the proportional increment rule.

In reality, the price tends to move up if there are more buy orders
than sell orders; similarly, the price tends to move down if there are
more sell orders than buy orders. The FI rule is meant to model the
sign but not the magnitude of the slope of this correlation, while the
PI rule attempts to model both. Clearly, there can be many other
increment rules, which this paper leaves for future research.

\section{Predicting the Market}
\label{section-prediction}

Informally, the \buzz{market prediction problem} at the beginning of
day $t$ is defined as follows:
\begin{itemize}
\item 
The data consists of (1) the five parameters of an AS-model, i.e.,
$m$, $\alpha$, $\Pi$, $X_i$, and a price adjustment rule, and (2) a
price history $P_0,\ldots,P_{t-1}$.
\item 
The goal is to predict the price $P_t$ by estimating the conditional
probabilities $\Pr[P_t>P_{t-1} \mid P_0,\ldots,P_{t-1}]$,
$\Pr[P_t<P_{t-1} \mid P_0,\ldots,P_{t-1}]$, and
$\Pr[P_t=P_{t-1} \mid P_0,\ldots,P_{t-1}]$.
\end{itemize}
Note that $\Pr[P_t>P_{t-1} \mid P_0,\ldots,P_{t-1}]$ is symmetric to
$\Pr[P_t<P_{t-1} \mid P_0,\ldots,P_{t-1}]$ and
$\Pr[P_t=P_{t-1} \mid P_0,\ldots,P_{t-1}] = 1 -
\Pr[P_t>P_{t-1} \mid P_0,\ldots,P_{t-1}]-\Pr[P_t<P_{t-1} \mid P_0,\ldots,P_{t-1}]$.
Thus, from this point onwards, our discussion focuses on estimating
$\Pr[P_t>P_{t-1} \mid P_0,\ldots,P_{t-1}]$.

From an algorithmic perspective, we sometimes assume that the price
adjustment rule and the joint distribution of the variables $X_i$ are
fixed, and that the input to the algorithm is $m$, $\alpha$, a
description of $\Pi$, and the price history.  This allows different
algorithms for different model families as well as side-steps the
issue of how to represent the possibly very complicated joint
distribution of the variables $X_i$ as part of the input. As for the
description of $\Pi$, we only need $\SSS^i_1,\ldots,\SSS^i_t$ for each $\SSS^i \in \Pi$ 
instead of
the whole $\Pi$, and the description of these functions can simplified
by restricting their domains to consist of the price sequences
consistent with the given price history.

\subsection{Markets as Systems of Linear Constraints}
\label{section-market-to-lp}

In the AS+FI model with parameters $m$ and $\alpha$, a price sequence
$P_0,\ldots,P_t$ and $\Pi$ can yield a set of linear inequalities in
the population variables $X_i$ as follows.  If the price changes on
day $t$, we have
\begin{equation}\label{eq-linear-change}
\sgn(P_t - P_{t-1}) \sum_{i=1}^h \SSS^i_t X_i > 0.
\end{equation}

If the price does not change, we have instead the equation
\begin{equation}\label{eq-linear-nochange}
\sum_{i=1}^h \SSS^i_t X_i = 0.
\end{equation}

Furthermore, any assignment of the variables $X_i$ that satisfies
either inequality is feasible with respect to the corresponding price
movement on day $t$.  In both cases, $\SSS^i_t$ is computable from the
price sequence $P_0,\ldots,P_{t-1}$.  The same statements hold for
days $1,\ldots,t-1$.  Therefore, given $m$ and $\alpha$, we can
extract from $\Pi$ and $P_0,\ldots,P_t$ a set of linear constraints on
the variables $X_i$. The converse holds similarly.  We formalize these
two observations in Lemmas~\ref{lemma-FI-market-to-inequalities} and
\ref{lemma-inequalities-to-FI-market} below.

\begin{lemma}
\label{lemma-FI-market-to-inequalities}
In the AS+FI model with parameters $m$ and $\alpha$, given $\Pi$ and a
price sequence $P_0,\ldots,P_{\beta}$, there are matrices $A$ and
$B$ with coefficients in $\sss$, $h$ columns each, and $\beta$ rows
in total. The rows of $A$ (respectively, $B$) correspond to the days
when $P_j \neq P_{j-1}$ (respectively, $P_j=P_{j-1}$).
Furthermore. the column vectors $x = (X_1,\ldots,X_h)^{\trans}$ consistent with $\Pi$
and $P_0,\ldots,P_{\beta}$ are exactly those that satisfy $Ax > 0$
and $Bx = 0$.  The matrices $A$ and $B$ can be computed in time
$O(h{\beta}T)$, where $T$ is an upper bound on the time to compute a
single $\SSS^i_j$ from $P_0,\ldots,P_{\beta}$ over all $j \in
[1,\beta]$ and $\SSS^i$.
\end{lemma}
\begin{proof}
Follows immediately from Equations~\ref{eq-linear-change}
and 
\ref{eq-linear-nochange}.
\end{proof}

\begin{lemma}\label{lemma-inequalities-to-FI-market}
In the AS+FI model with parameters $m$ and $\alpha$, given a system of
linear inequalities $Ax > 0, Bx = 0$, where $A$ and $B$ have
coefficients in $\sss$ with $h$ columns each, and $\beta$ rows in
total, there exist $(1)$ a set $\Pi$ of $h$ strategies corresponding
to the $h$ columns of $A$ and $B$, and $(2)$ a $(\beta+1)$-day price
sequence $P_0,\ldots,P_\beta$ with the latter $\beta$ days
corresponding to the $\beta$ rows of $A$ and $B$. Furthermore, the
values of the population variables $X_1,\ldots,X_n$ are feasible
with respect to the price movement on day $j$ if and only if column vector $x =
(X_1,\ldots,X_n)^{\trans}$ satisfies the $j$-th constraint in $A$ and $B$.
Also, $P_0,\ldots,P_\beta$ and a description of $\Pi$ can be computed
in $O(h\beta)$ time.
\end{lemma}
\begin{proof}
Follows immediately from Equations~\ref{eq-linear-change}
and 
\ref{eq-linear-nochange}.
\end{proof}

In the AS+PI model we obtain only equations, of the form:
\begin{equation}
\label{eq-linear-proportional}
\sum_{i=1}^h \SSS^i_t X_i = \frac{1}{\alpha} (P_t - P_{t-1}).
\end{equation}
In this case there is a direct correspondence between market
data and systems of linear equations.  We formalize this
correspondence in Lemmas~\ref{lemma-PI-market-to-equations} and
\ref{lemma-equations-to-PI-market} below.

\begin{lemma}
\label{lemma-PI-market-to-equations}
In the AS+PI model with parameters $m$ and $\alpha$, given $\Pi$ and a
price sequence $P_0,\ldots,P_{\beta}$, there is a matrix $B$ 
with coefficients in $\sss$, $h$ columns, and $\beta$ rows, and a
column vector $b$ of length $h$, such that 
the column vectors $x = (X_1,\ldots,X_h)^{\trans}$ consistent with $\Pi$
and $P_0,\ldots,P_{\beta}$ are exactly those that satisfy $Bx = b$.
The coefficients of $B$ and $b$ can be computed in time
$O(h{\beta}T)$, where $T$ is an upper bound on the time to compute a
single $\SSS^i_j$ from $P_0,\ldots,P_{\beta}$ over all $j \in
[1,\beta]$ and $\SSS^i$.
\end{lemma}
\begin{proof}
Follows immediately from Equation \ref{eq-linear-proportional}.
\end{proof}

\begin{lemma}\label{lemma-equations-to-PI-market}
In the AS+PI model with parameters $m$ and $\alpha$, given a system of
linear equations $Bx = b$, where $B$ is a $\beta \times h$ matrix with
coefficients in $\sss$, 
there exist $(1)$ a set $\Pi$ of $h$ strategies corresponding
to the $h$ columns of $B$, and $(2)$ a $(\beta+1)$-day price
sequence $P_0,\ldots,P_\beta$ with the last $\beta$ days
corresponding to the $\beta$ rows of $B$. Furthermore, the
values of the population variables $X_1,\ldots,X_n$ are feasible
with respect to the price movement on day $j$ if and only if column vector $x =
(X_1,\ldots,X_n)^{\trans}$ satisfies the $j$-th constraint in $B$.
Also, $P_0,\ldots,P_\beta$ and a description of $\Pi$ can be computed
in $O(h\beta)$ time.
\end{lemma}
\begin{proof}
Follows immediately from Equation \ref{eq-linear-proportional}.
\end{proof}

\subsection{An Easy Case for Market Prediction: Many Traders but Few Strategies}
\label{section-prediction-easy}
In Section~\ref{section-prediction-easy-FI}, we show that if an AS+FI market
has far more traders than strategies, then it takes polynomial
time to estimate the probability that the next day's price will rise.
In Section \ref{section-prediction-easy-PI}, we discuss why the same
analysis technique does not work for an AS+PI market.

\subsubsection{Predicting an AS+FI Market}
\label{section-prediction-easy-FI}
For the sake of emphasizing the dependence on $m$, let $\Prm[E]$ be
the probability that event $E$ occurs when there are $m$ traders in
the market.

\newcounter{easy}

This section makes the following assumptions:
\begin{list}{\theeasy}
{\usecounter{easy}\setcounter{easy}{0}\renewcommand{\theeasy}{E\arabic{easy}}
\setlength{\rightmargin}{0in} 
\settowidth{\leftmargin}{E4} \addtolength{\leftmargin}{\labelsep}}
\item\label{e1} 
The input to the market prediction problem is simply a price history
$P_0,\ldots,P_{t-1}$.  The output is $\limm\Prm[P_t > P_{t-1} \mid 
P_0,\ldots,P_{t-1}]$.
\item\label{e2}  
The market follows the AS+FI model.
\item\label{e3}   
$\Pi$ is fixed.  The values $\SSS^i_j$ over all $i \in [1,h]$ are
computable from the input in total time polynomial in $j$.
\item\label{e4}   
Each of the $m$ traders independently chooses a random strategy
$\SSS^i$ from $\Pi$ with fixed probability $p_i > 0$, where
$p_1+\cdots+p_h = 1$.
\end{list}
The parameter $\alpha$ is irrelevant.

Notice that the column vector $X=(X_1,\ldots,X_h)^{\trans}$ is the sum
of $m$ independent identically-distributed vector-valued random
variables with a center at $p=m{\cdot}(p_1,\ldots,p_h)^{\trans}$.  We
recenter and rescale $X$ to $\overlineX =
(X-m{\cdot}(p1,\ldots,p_h)^{\trans})/\sqrt{m}$. Then, by the Central
Limit Theorem (see, e.g., \cite[Theorem 29.5]{Billingsley}), as $m
\goesto +\infty$, $\overlineX$ converges weakly to a normal
distribution centered at
the $h$-dimensional vector
$(0,\ldots,0)^{\trans}$. In
Theorem~\ref{theorem-prediction-easy-fixed} below, we rely on this
fact to estimate the probability that the market rises for price
histories that occur with nonzero probability.

\begin{theorem}
\label{theorem-prediction-easy-fixed}
Assume that $\limm\Prm[P_0,\ldots,P_{t-1}]>0$. Then there is a
fully polynomial-time approximation scheme for estimating
$\limm\Prm[P_t > P_{t-1} \mid P_0,\ldots,P_{t-1}]$ from
$P_0,\ldots,P_{t-1}$.  The time complexity of the scheme is polynomial
in $(1)$ the length $t$ of the price history, 
$(2)$ the inverse of the relative error bound
$\epsilon$, and $(3)$ the inverse of the failure probability $\eta$.
\end{theorem}
{\it Remark.}  We omit the explicit dependency of the running time in
$h$ and $p_1,\ldots,p_h$ in order to concentrate on the main point
that market prediction is easy with this section's four assumptions.  The
parameters $h$ and $p_1,\ldots,p_h$ are constant under the
assumptions.
\begin{proof}
We use Lemma~\ref{lemma-FI-market-to-inequalities} to convert the price
history $P_0,\ldots,P_{t-1}$ and the strategy set $\Pi$ into a system
of linear constraints $AX > 0$ and $BX=0$, with the next day's price
change $P_t-P_{t-1}$ determined by $\sgn(c{\cdot}X')$ for some $c$.
Since the values $\SSS^i_j$ are computable in time polynomial in $j$,
this conversion takes time polynomial in $t$.

Then, $\Prm[P_0,\ldots,P_{t-1}]=\Prm[AX>0 \wedge BX=0]$. Since
$\limm\Prm[AX > 0 \wedge BX=0] > 0$, the constraints in $B$ must be
vacuous; in other words, for each $P_i=0$ with $i\in[0,t-1]$, the
corresponding constraint in $B$ is
$0{\cdot}X_1+\cdots+0{\cdot}X_h=0$.
Therefore,
$\Prm[P_0,\ldots,P_{t-1}]=\Prm[AX>0]$.  Furthermore, since both
$A$ and $c$ are constant with respect to $m$,
\begin{equation}
\label{eqn_ratio_1}
\limm\Prm[P_t > P_{t-1} \mid P_0,\ldots,P_{t-1}] 
= 
\frac{\limm\Prm[AX > 0 \wedge c{\cdot}X > 0]}{\limm\Prm[AX > 0]}.
\end{equation} 
So to compute the desired $\limm\Prm[P_t > P_{t-1} \mid 
P_0,\ldots,P_{t-1}]$, we compute $\limm\Prm[AX > 0 \wedge c{\cdot}X
> 0]$ and $\limm\Prm[AX > 0]$ as follows.

To avoid the degeneracy caused by $\sum_{i=1}^{h} X_i = m$, we work
with $X'=(X_1,\ldots,X_{h-1})^{\trans}$ instead of $X$ by replacing
$X_h$ with $m-\sum_{i=1}^{h-1} X_i$ and making related changes.  Let
$p'=(p_1,\ldots,p_{h-1})^{\trans}$, which is the center of $X'$.  As
is true for $\overlineX$, as $m \goesto +\infty$, the vector
$\overlineX'=(X'-m{\cdot}p')/\sqrt{m}$ converges weakly to a normal
distribution centered at the $(h-1)$-dimensional point
$(0,\ldots,0)^{\trans}$ .  
Under the assumption that each $p_i$ is nonzero,
the distribution of $Y'$ is full-dimensional
(within its restricted $(h-1)$-dimensional space),
as in the limit the variance of each coordinate $Y'_i$ is nonzero
conditioned on the values of the other coordinates,
which implies that the smallest subspace containing the distribution
must contain all $h-1$ axes.
We can 
calculate the covariance matrix of $Y'$ directly from the $p_i$, as it
is equal to the covariance matrix for a single trader:
on the diagonal,
$C_{ii} = p_i - p_i^2$;
and for off-diagonal elements,
$C_{ij} = -p_i p_j$.
Given $C$, $Y'$ has density $\rho(x) = a e^{x^\trans C x}$
for some constant $a$, and we can evaluate this density in $O(h^2)$ time
given $x$, which is $O(1)$ time under our assumption that $\Pi$ is
fixed.

Let
$A_i$ be the $i$-th constraint of $A$, i.e.,
$A_{i,1}X_1+\cdots+A_{i,h}X_h > 0$. Let $A'_i$ denote the constraint
$(A_{i,1}-A_{i,h},\ldots,A_{i,{h-1}}-A_{i,h})$.  Let
$c'=(c_1-c_h,\ldots,c_{h-1}-c_h)$.

We next convert the constraints of $A$ on $X$ into constraints on
$\overlineX'$.  First of all, notice that $A_i X =
\sqrt{m}{\cdot}(A'_i \overlineX') + m{\cdot}A_i p$.  So $A_i X > 0$ if
and only if $A'_i \overlineX' > -\sqrt{m}{\cdot}A_i p$.  The term
$-\sqrt{m}{\cdot}A_i p$ may not be constant. In such a case, as $m
\goesto \infty$, the hyper plane bounding the half space $A'_i
\overlineX' > -\sqrt{m}{\cdot}A_i p$ keeps moving away from the
origin, which presents some technical complication.  To remove this
problem, we analyze the term in three cases.  If $A_i p < 0$, then
since $m{\cdot}p$ is the center of $X$, as $m \goesto \infty$,
$\Prm[A_i X<0]$ converges to $1$.  In other words, $A_i$ is infeasible
with probability $1$ in the limit. Then, since
$\limm\Prm[P_0,\ldots,P_{t-1}]>0$, such $A_i$ cannot exist in $A$.
Similarly, if $A_i p > 0$, then $\limm\Prm[A_i X > 0]=1$ and $A_i$ is
vacuous.  The interesting constraints are those for which $A_i p = 0$;
in this case, by algebra, $A_i X > 0$ if and only if $A'_i \overlineX'
> 0$.  Thus, let $\overlineA$ be the matrix formed by these
constraints; $D$ can be computed in $O(ht)$ time.  Then, since $D$ is
constant with respect to $m$, $\limm\Prm[AX >
0]=\limm\Prm[\overlineA\overlineX'>0]$.  Similarly, $\Prm[AX > 0
\wedge c{\cdot}X > 0]$ converges to (1) $0$, (2)
$\Prm[\overlineA\overlineX' > 0]$, or (3) $\Prm[\overlineA\overlineX'
> 0 \wedge c'{\cdot}\overlineX' > 0]$ for case (1) $c{\cdot}p < 0$,
case (2) $c{\cdot}p > 0$, or case (3) $c{\cdot}p = 0$, respectively.

Therefore, by Equation~\ref{eqn_ratio_1}, $\limm\Prm[P_t > P_{t-1} \mid 
P_0,\ldots,P_{t-1}]$ equals $0$ for case (1) and equals $1$ for case
(2).  Case (3) requires further computation.
\begin{equation}
\label{eqn_ratio}
\limm\Prm[P_t > P_{t-1} \mid P_0,\ldots,P_{t-1}] 
=
\frac{\limm\Prm[\overlineA\overlineX' > 0 \wedge 
c'{\cdot}\overlineX' > 0]}{\limm\Prm[\overlineA\overlineX' > 0]}.
\end{equation}
The numerator and denominator of the ratio in Equation~\ref{eqn_ratio}
are both integrals of the distribution of $\overlineX'$ in the limit
over the bodies of possibly infinite convex polytopes.  
To deal with the
possible infiniteness of the convex bodies $\overlineA\overlineX' > 0
\wedge c'{\cdot}\overlineX' > 0$ and $\overlineA\overlineX' > 0$,
notice that the density drops exponentially. So we can truncate the
regions of integration to some finite radius around the
$(h-1)$-dimensional origin $(0,\ldots,0)^{\trans}$ with only
exponentially small loss of precision.  
Finally, since 
the distribution of $\overlineX'$ in the limit is normal, 
by applying the
Applegate-Kannan integration algorithm for log-concave distributions
\cite{ApplegateK91} to the numerator and denominator separately, we
can approximate $\limm\Prm[P_t > P_{t-1} \mid P_0,\ldots,P_{t-1}]$ within
the desired time complexity.
\end{proof}

\subsubsection{Remarks on Predicting an AS+PI Market}
\label{section-prediction-easy-PI}
The probability estimation technique based on taking $m$ to $\infty$
does not appear to be applicable to the AS+PI model for the following
reasons.

First of all, by Lemma~\ref{lemma-PI-market-to-equations}, the input
price history induces a system of linear equations $BX=b$.  If any
equation in $BX=b$ is not equivalent to $X_1+\cdots+X_h=m$ or
$0{\cdot}X_1+\cdots+0{\cdot}X_h=0$, then
$\limm\Prm[P_0,\ldots,P_{t-1}]=0$.

A natural attempt to overcome this seemingly technical difficulty
would be to (1) solve $BX=b$ to choose a maximal set $U$ of
independent variables $X_i$ and (2) evaluate
$\Prm[P_0,\ldots,P_{t-1}]$ in the probability space induced by this
set. Still, a single constraint such as
$B_{i,1}{\cdot}X_1+\cdots+B_{i,h}{\cdot}X_{h}=\alpha{\cdot}m_0$ with
$B_{i,j} \geq 0$ for all $j \in [1,h]$ and $B_{i,j'} > 0$ for some
$X_{j'} \in U$ forces $\limm\Prm[P_0,\ldots,P_{t-1}]=0$ in the new
probability space.  This is due to the fact that $m_0$ is constant
with respect to $m$.

A further attempt would be to evaluate $\limm\Prm[P_t > P_{t-1} \mid 
P_0,\ldots,P_{t-1}]$ by directly working with the probability space
induced by $P_0,\ldots,P_{t-1}$.  This also does not work because we
show below that the market prediction problem can be reduced to the
case where taking a limit in $m$ has no effect on the distribution of
the strategy counts.  Suppose that we are given a market which
follows the assumptions \ref{e1}, \ref{e3}, and \ref{e4} of
Section \ref{section-prediction-easy-FI} except that this market uses the PI
rule and has $m_0$ traders.  We construct a new market with any $m
\geq m_0$ traders with the following modifications:
\begin{enumerate}
\item 
The price history $P_0,\ldots,P_{t-1}$ is extended with an extra day
into $P'_0,\ldots,P'_{t-1},P'_t$, where $P'_j=P_j$ for $0 \leq j \leq
t-1$.  Each strategy $\SSS_i$ is extended into a new strategy
$\SSS'_i$ where (1) on day $j\in[1,t-1]$,
$\SSS'_i(P_0,\ldots,P_{j-1})=\SSS_i(P_0,\ldots,P_{j-1})$, (2) on day
$t$, $\SSS'_i$ always buys, and (3) on day $t+1$,
$\SSS'_i(P'_0,\ldots,P'_t)=\SSS_i(P_0,\ldots,P_{t-1})$.  Thus,
$P'_t=P'_{t-1}+\alpha{\cdot}m_0$.
\item 
Add a passive strategy $\SSS'_{h+1}$ that always holds.
\item 
Let $p'_i = \frac{1}{2}p_i$ for $1 \le i \le h$ and
$p'_{h+1}=\frac{1}{2}$.
\end{enumerate}
Note that since $P'_t-P'_{t-1}=\alpha{\cdot}m_0$, $m-m_0$ traders
choose the passive strategy $\SSS_{h+1}$. Also, the new market and the
new price history can accommodate any $m \geq m_0$ traders. Note that
because of the constraint $P'_{t}-P'_{t-1}=\alpha{\cdot}m_0$, the
probability distribution of $(X_1,\cdots,X_h)^{\trans}$ conditioned on
$P'_0,\ldots,P'_t$ in the new market for each $m \geq m_0$ is
identical to the probability distribution of
$(X_1,\cdots,X_h)^{\trans}$ conditioned on $P_0,\ldots,P_{t-1}$ in the
original market with $m=m_0$.  Furthermore, $\Prm[P'_{t+1} >
P'_t \mid P'_0,\ldots,P'_t] = {\rm
Pr}_{m_0}[P_t>P_{t-1} \mid P_0,\ldots,P_{t-1}]$.  So we have obtained the
desired reduction.

Consequently, we are left with a situation where the number of active
strategies may be comparable to the number of traders.  Such a market
turns out to be very hard to predict, as shown next in
Section~\ref{section-prediction-hard}.

\subsection{A Hard Case for Market Prediction: Many Strategies}
\label{section-prediction-hard}
Section~\ref{section-prediction-easy} shows that predicting an AS+FI market
is easy (i.e., takes polynomial time) when the number $m$ of traders
vastly exceeds the number $h$ of strategies.  In this section, we
consider the case where every trader may have a distinct strategy, and
show that predicting an AS+FI or AS+PI market becomes very hard indeed.

We now define two decision-problem versions of market prediction. Both
versions make the following assumption:
\begin{itemize}
\item 
Each $X_i$ is independently either $0$ or $1$ with equal probability.
\end{itemize}

The \buzz{bounded} market prediction problem is:
\begin{itemize}
\item Input: a set of $n$ passive strategies and a price history
spanning $n$ days such that the probability that the market rises on
day $n+1$ conditioned on the price history is either (1) greater than
$2/3$ or (2) less than $1/3$.
\item Question: Which case is it, case (1) or case (2)?
\end{itemize}

The \buzz{unbounded} market prediction problem is:
\begin{itemize}
\item 
Input: a set of $n$ passive strategies and a price history spanning
$n$ days.
\item 
Question: Is the probability that the market rises on day $n+1$
conditioned on the price history greater than 1/2 (without the usual
$\epsilon$ term)?
\end{itemize}

The unbounded market prediction problem has less financial payoff than
the bounded one due to different probability thresholds.  For each of
these two problems, there are in effect two versions, depending on
which price increment rule is used; however, both versions turn out to
be equally hard. These two problems can be analyzed by similar
techniques, and our discussion below focuses on the bounded market
prediction problem with a hardness theorem for the unbounded market
prediction problem in Section~\ref{sec_unbounded}.

We show in Section \ref{section-circuit-to-market} how to construct
passive strategies and price histories such that solving bounded
market prediction is equivalent to estimating the probability that a
Boolean circuit outputs $1$ on a random input conditioned on a second
circuit outputting $1$.  In Section \ref{section-bcpp}, we show that
this problem is hard for $\pnplog$ and complete for a class that lies
between $\pnplog$ and $\pp$.  Thus bounded market prediction is not
merely \np-hard, but cannot be solved in the polynomial-time hierarchy
at all unless the hierarchy collapses to a finite level.

\subsubsection{Reductions from Circuits to Markets}
\label{section-circuit-to-market}

Lemma~\ref{lemma-circuit-to-inequalities} converts a
circuit into a system of linear inequalities, while
Lemma~\ref{lemma-slack-variables} converts a system of
linear inequalities into a system of linear equations.
These systems can then be converted into AS+FI and AS+PI market models
using Lemmas~\ref{lemma-inequalities-to-FI-market} and
\ref{lemma-equations-to-PI-market}, respectively.

Note that the restriction in Lemma~\ref{lemma-circuit-to-inequalities}
to circuits consisting of $2$-input NOR gates is not an obstacle to
representing arbitrary combinatorial circuits (with constant blow-up),
as $2$-input NOR gates are universal.

\begin{lemma}
\label{lemma-circuit-to-inequalities}
For any $n$-input Boolean circuit $C$
consisting of $m$ 2-input NOR gates,
there exists a system $Ax > 0$ of $3m+2$ linear constraints in $n+m+2$ unknowns
and a length $n+m+2$
column vector $c$
with the following properties:
\begin{enumerate}
\item Both $A$ and $c$ have coefficients in $\sss$ that can be
computed in time $O((n+m)^2)$.
\item 
Any $0$-$1$ vector $(x_1,\ldots,x_n)$
has a unique $0$-$1$ extension 
$x = (x_1,\ldots,x_n, x_{n+1}, \ldots x_{n+m+2})$
satisfying $Ax > 0$.
\item 
If $Ax > 0$, then $cx > 0$ if and only if
$C(x_1, x_2,\ldots,x_n) = 1$.
\end{enumerate}
\end{lemma}
\begin{proof}
Let $x_{n+k}$ represent the output of the $k$-th NOR gate, where $1
\le k \le m$.  Without loss of generality we assume that gate $m$ is
the output gate.

The variables $x_{n+m+1}$ and $x_{n+m+2}$ are dummies to allow for a
zero right-hand-side in $Ax > 0$; our first two constraints
are $x_{n+m+1} > 0$ and $x_{n+m+2} > 0$.

Suppose gate $k$ has inputs $x_i$ and $x_j$.  
The NOR operation is
implemented by the following three linear inequalities:
\begin{displaymath}
\begin{array}{ccccccc}
x_i&  &   & + &x_{n+k} & <& 2; \\
   & &x_j& + &x_{n+k} & < &2; \\
x_i &+&x_j&+&x_{n+k}& >& 0.    	
\end{array}
\end{displaymath}
The first two constraints
ensure that the output is never $1$ if an input is $1$,
while the last requires that the output is $1$ if both inputs are $0$;
the constraints are thus satisfied if and only if 
$x_{n+k} = \neg(x_i \vee x_j)$.
Using the dummy variables, the first two constraints are written as
\begin{displaymath}
\begin{array}{ccccccccccc}
-x_i&  &    & - &x_{n+k} &+& x_{n+m+1} &+& x_{n+m+2} & >& 0; \\
    &  &-x_j& - &x_{n+k} &+& x_{n+m+1} &+& x_{n+m+2} & >& 0. \\
\end{array}
\end{displaymath}

Let $Ax > 0$ be the system obtained by combining all of these inequalities.
Then for each $(x_1,\ldots,x_n)$, $Ax > 0$ determines $x_{n+k}$ for all 
$k \ge 1$.
The vector $c$ is chosen so that $cx = x_{n+m}$.
\end{proof}

One might suspect that the fixed increment rule's ability to
hide the exact values of the left-hand side of each constraint is
critical to disguise the inner workings of the circuit.  
However, by adding slack variables we can translate the inequalities
into equations, allowing the use of a proportional increment rule without
revealing extra information.

\newcommand{\lemmaSlackEquations}{mn - m + 1}
\newcommand{\lemmaSlackVariables}{2mn - 3m + n + 1}
\begin{lemma}
\label{lemma-slack-variables}
Let $Ax > 0$ be a system of $m$
linear inequalities in $n$ variables where $A$ has
coefficients in $\sss$.  
Then there is a system $By = 1$ of $\lemmaSlackEquations$ linear equations
in $\lemmaSlackVariables$ variables
with the following properties:
\begin{enumerate}
\item $B$ has coefficients in $\sss$ that can be computed in
time $O((mn)^2)$.
\item There is a bijection $f : x \mapsto y$ between the $0$-$1$
solutions $x$ to
$Ax > 0$ and 
the $0$-$1$ solutions
$y$ to $By = 1$,
such that
$x_j = y_j$ for $1 \le j \le n$ whenever $y = f(x)$.
\end{enumerate}
\end{lemma}
\begin{proof}
For each $1 \le i \le m$, let $A_i$ be the constraint
$\sum_j A_{ij} x_j > 0$.
To turn these inequalities into equations, we
add slack variables to soak up
any excess over $1$, with some
additional care taken to ensure that there is a unique assignment to
the slack variables for each setting of the variables $x_j$.

We will use the following $0$-$1$ variables, which we think of as
alternate names for $y_1$ through $y_{\lemmaSlackVariables}$:

\begin{center}
\begin{tabular}{|c|l|c|c|}
\hline
Variables & Purpose & Indices & Count  \\
\hline
$x_j$ & original variables & $1 \le j \le n$ & n \\
$u$    & constant $1$    &  none     & 1 \\
$s_{ij}$  & slack variables for $A_i$ & $1 \le i \le m, 1 \le j \le
n-1$ & $m(n-1)$ \\
$t_{ij}$  & slack variables for $s_{ij} \ge s_{i,j+1}$ 
    & $1 \le i \le m, 1 \le j \le n-2$ & $m(n-2)$ \\
\hline
\end{tabular}
\end{center}

\begin{center}
\begin{tabular}{|c|c|l|c|c|}
\hline
Name & Equation & Purpose & Indices & Count \\
\hline
$U$ & $u = 1$ & set $u$ & none & 1 \\
$B_i$ & $\sum_j A_{ij} x_j - \sum_j s_{ij} = 1$
   & represent $A_i$ & $1 \le i \le m$ & $m$ \\
$S_{ij}$ & $s_{ij} - s_{i,j+1} - t_{ij} + u = 1$ 
   & require $s_{ij} \ge s_{i,j+1}$ & $1 \le i \le m$, $1 \le j \le n-2$
   & $m(n-2)$
\\
\hline
\end{tabular}
\end{center}

Observe that for each $i$, $\sum_j s_{ij}$ can take on any integer
value $\sigma_i$ between $0$ and $n-1$, and that for any fixed value of
$\sigma_i$, the $S_{ij}$ constraints uniquely determine the values of
$s_{ij}$ and $t_{ij}$ for all $j$.  So each 
constraint $B_i$ permits 
$\chi_{i} = \sum_j A_{ij} x_j$ 
to take on precisely the same values $1$ to $n$
that $A_i$ does,
and each
$\chi_i$ uniquely determines $\sigma_i$ and thus the assignment of all
$s_{ij}$ and $t_{ij}$.
\end{proof}

\subsubsection{Conditional Probability Complexity Classes}
\label{section-bcpp}
Suppose that we take a polynomial-time probabilistic Turing machine, fix its
inputs, and use the usual Cook's Theorem construction to turn it into
a circuit whose inputs are the random bits used during its
computation. Then, we can feed the resulting circuit to
Lemmas~\ref{lemma-circuit-to-inequalities} and
\ref{lemma-inequalities-to-FI-market} 
to obtain  an AS+FI market model in which there is exactly one assignment of
population variables for each set of random bits, and the price rises
on the last day if and only if the output of the Turing machine is
$1$.  
By applying Lemma~\ref{lemma-slack-variables} to the intermediate system
of linear inequalities, we can similarly convert a circuit to an AS+PI
model.
It follows that bounded market prediction is \bpp-hard for either
model.
But with some cleverness,
we can exploit the conditioning on past history to show that bounded
market prediction is in fact much harder than this.  We do so in
Section~\ref{section-bcpp-complete}, after a brief detour through
computational complexity in this section.

We proceed to define some new counting classes based on
conditional probabilities.  One of these, \bcpp{}, has the useful
feature that bounded market prediction is \bcpp-complete.  We will use
this fact to relate the complexity of bounded market prediction to
more traditional complexity classes.

The usual counting classes of complexity theory ($\pp{}$, $\bpp{}$, R, $\zpp{}$,
$C_{=}$, etc.) are defined in terms of counting the relative numbers of
accepting and rejecting states of a nondeterministic Turing machine.
We will define a new family of counting classes
by adding a third decision state that does not count for the purposes
of determining acceptance or rejection.

A {\it noncommittal} Turing machine is a nondeterministic Turing
machine with three decision states: {\it accept}, {\it reject}, and {\it abstain}.  
We 
represent a noncommittal Turing machine as a deterministic Turing
machine which takes a polynomial number of random bits in addition to
its input; each assignment of the random bits gives a distinct
computation path.  A computation path is
accepting/rejecting/abstaining if it ends in an accept/reject/abstain
state, respectively.  We often write $1$, $0$, or $\perp$ as shorthand for the
output of an accepting, rejecting, or abstaining path.

Conditional versions of the usual counting classes are obtained by
carrying over their definitions from standard nondeterministic Turing
machines to noncommittal Turing machines, with some care in handling
the case of no accepting or rejecting paths.  We can still think of
these modified classes as corresponding to probabilistic machines, but
now the probabilities we are interested in are conditioned on not
abstaining.

\begin{definition}\label{def-cpp}\rm
The {\it conditional probabilistic polynomial-time} class (\cpp{})
consists of those languages $L$ for which there exists a
polynomial-time noncommittal Turing machine $M$ such that $x \in L$ if
and only if the number of accepting paths when $M$ is run with input
$x$ exceeds the number of rejecting paths.
\end{definition}

\begin{definition}\label{def-bcpp}\rm
The {\it bounded conditional probabilistic polynomial-time} class
(\bcpp{}) consists of those languages $L$ for which there exists a
constant $\epsilon > 0$ and a polynomial-time noncommittal Turing
machine $M$ such that (1) $x \in L$ implies that a fraction of at
least $\frac{1}{2}+\epsilon$ of the total number of accepting and
rejecting paths are accepting and (2) $x \notin L$ implies that a
fraction of at least $\frac{1}{2}+\epsilon$ of the total number of
accepting and rejecting paths are rejecting.
\end{definition}

\begin{definition}\label{def-cr}\rm
The \buzz{conditional randomized polynomial-time} class (\crclass{})
consists of those languages $L$ for which there exists
a constant $\epsilon > 0$ and a
polynomial-time noncommittal
Turing machine $M$ such that (1) $x \in L$
implies that a fraction of at least $\epsilon$ 
of the total number of accepting and rejecting paths are accepting,
and (2) $x \notin L$
implies that 
there are no accepting paths.
\end{definition}

As we show in Theorems~\ref{theorem-cpp-subset-pp}
and~\ref{theorem-cr-equals-np},
\cpp{} and \crclass{} turn out to be the same as the unconditional
classes \pp{} and \np{}, respectively.

\begin{theorem}
\label{theorem-cpp-subset-pp}
$\cpp{}=\pp{}$.
\end{theorem}
\begin{proof}
First of all, $\pp{} \subseteq \cpp{}$ because a $\pp{}$ machine is a \cpp{}
machine that happens not to have any abstaining paths.
For the inverse direction, represent each abstaining path of a \cpp{}
machine by a pair consisting of one accepting and one rejecting path, 
and each
accepting or rejecting path by two accepting or rejecting paths.  
Then the resulting $\pp{}$ machine accepts if and only if the
\cpp{} machine does.
\end{proof}

\begin{theorem}
\label{theorem-cr-equals-np}
$\crclass = \np$.
\end{theorem}
\begin{proof}
To show $\np \subseteq \crclass$, replace each
rejecting path of an
$\np$ machine with an abstaining path in a $\crclass$ machine.
For the inverse direction,
replace each abstaining path of the $\crclass$ machine 
with a rejecting path in the $\np$ machine.
\end{proof}

\bcpp{} appears to be a more interesting class.  Since it is clearly a
subset of \cpp{}, we have:

\begin{corollary}
\label{corollary-bcpp-subset-pp}
$\bcpp{} \subseteq \pp{}$.
\end{corollary}
\begin{proof}
Immediate from Theorem~\ref{theorem-cpp-subset-pp} and the definition
of \bcpp{} and \cpp{}.
\end{proof}

On the other hand, $\bcpp$ appears to be much stronger than the
analogous non-conditional class $\bpp$.  For example, it is straightforward
to show that $\np \subseteq \bpp$.  
Use the representation of an \np-machine as a deterministic
machine $M$ that takes some polynomial number of ``hint'' bits in addition
to its input, and replace these $N$ hint bits with $N$ random bits $r$.  In
addition, supply another $2N$ random bits $r'$, which will be used to
amplify the conditional probability of accepting paths.  Now let
$M'(x, r, r')$ accept if $M(x, r)$ accepts; reject if $M(x,r)$
rejects and $r'= \vec{0}$; and abstain if $M(x,r)$ rejects and
$r'\neq\vec{0}$.  Then if $M$ has any accepting path on input $x$, 
$M'$ has at
least $2^{2N}$ accepting paths and at most $2^N-1$ rejecting paths,
for an exponentially large probability of accepting--- since we have
amplified the small number of accepting paths so that they overwhelm
the few rejectors.
Alternatively, if $M(x,r)$ never accepts, neither does $M'$.

By repeating this sort of amplification of ``good'' paths, we can in fact
simulate $O(\log n)$ queries of an \np-oracle, as stated in the following theorem.

\begin{theorem}
\label{theorem-pnplog-subset-bcpp}
$\pnplog \subseteq \bcpp$.
\end{theorem}
\begin{proof}
Let $M(x,h_1,h_2,\ldots,h_k)$ be a deterministic
implementation a $\pnplog$ machine, 
where $k = O(\log n)$ and each $h_i$ supplies a witness for the 
$i$-th oracle query.  
We will show that the language $L(M)$ accepted by $M$ is in $\bcpp$.

To simplify the presentation, we assume that each oracle query is a
Boolean formula with a fixed number $m$ of variables, where $m$ is
polynomial in $n = |x|$, and that $h_i$ is an assignment for those
variables.  We assume that $M$ consists of a sequence of functions
$M_1,M_2,\ldots,M_k$ for generating oracle queries, a set of $k$
verifiers $V_1,\ldots,V_k$ for verifying the witnesses $h_i$, and a
combining function $M_*$ that produces the output from the input and
the outputs of the $V_i$.  Each function $M_i$ takes as input the
input $|x|$ and the outputs of $V_1$ through $V_{i-1}$. $V_i$ sees the
output of $M_i$ and the input $h_i$. $M_*$ sees the input $|x|$
and the outputs of $V_1$ through $V_k$.  The output of the combined
machine $M$ is the output of any computation path where $h_i$ is
chosen so that $V_i(M_i, h_i) = 1$ whenever such an $h_i$ exists.  In
other words, we demand that $h_i$ be a satisfying assignment when
possible, and ignore those paths where satisfiable queries are issued
but satisfying assignments are not supplied.

We will represent $M$ with a noncommittal
machine $M'(x,r_1,\ldots,r_k,r')$, where 
each $r_i$ is a random bit-vector replacing the corresponding $h_i$,
and $r'$ is an extra supply of random bits used to amplify the good
computation paths to overwhelm the bad ones.  This amplification
process is a little complicated, because it is not enough to amplify
paths that find good witnesses to particular queries; it may be that
a bad witness for an earlier query causes some $M_i$ to issue a different
query from the correct one.  So we must amplify a path that finds a
witness to query $i$ 
enough to overwhelm not only the exponentially many invalid
witnesses to query $i$, but also
the exponentially many valid witnesses that might be returned to
instances
queries $i+1$ through $k$ based on an incorrect answer to query $i$.

Let $v = (v_1,v_2,\ldots,v_k)$ be the vector of outputs of $V_i$.  For
each $v$, we define an \buzz{amplification exponent} $A(v)$ as follows:
\begin{displaymath}
A(v) = \sum_{i=1}^{k} v_i 2^{k-i}(mk+1).
\end{displaymath}
We will write $A_i$ for the coefficient $2^{(k-i)}(mk+1)$;
these coefficients $A_i$ are chosen to make Equation~\ref{eq-Ai-recurrence} work below.

Now let $M'(x,r_1,\ldots,r_k,r') = M(x,r_1,\ldots,r_k)$ whenever
$r'_i = 0$ for all $i > A(v)$, where $v$ is the output of
$V_1$ through $V_k$ in the computation of $M$.  If $r'_i \neq 0$ for
some $i > A(v)$, $M'(x,r_1,\ldots,r_k,r') = \perp$.
The effect of the $r'$ bits is to set the weight of
each non-abstaining path to $2^{A(v)}$.

Clearly $M'$ can be computed in polynomial time as long as $|r'|$ is
polynomial.
The number of $r'$ bits needed is the maximum value of $A(v)$,
which is $(mk+1) (2^{k+1} - 1) = O(n^c k 2^h)$.
For this to be polynomial, we need $k = O(\log n)$.

A good path $p$ is precisely one for which each
$v_i$ is the correct output of the \np-oracle.  A bad path $p'$ is one in
which one or more of the $v'_i$ values is incorrect.  
We will match bad paths with good paths, and show that the weight of
each good path is much larger than the total weight of all bad paths
mapped to it.

Identify each path with the sequence $r_1,\ldots,r_k$ that generates
it.  Let $b = b_1,\ldots,b_k$ generate some bad path.  Let $i$ be the
first point at which $b_i$ is an invalid witness to a satisfiable
query.  Then there is a good path $a = a_1,\ldots,a_k$ such that 
$a_j = b_j$ for $j < i$.  Furthermore, if $v^a$ and $v^b$ are the
vectors of verifier outputs for $a$ and $b$, then not only is 
$v^a_j = v^b_j$ for $j < i$, but also $v^a_i = 1$ while $v^b_i = 0$ since
the only false verifier outputs are false negatives.

The maximum value for $A(v^b)$ is obtained if $v^b_j = 1$ for $j > i$;
so we have
\begin{displaymath}
A(v_a) - A(v_b) 
\ge 
\left(\sum_{j=1}^{i-1} A_j v^a_j + A_i\right)
-
\left(\sum_{j=1}^{i-1} A_j v^b_j + \sum_{j=i+1}^{k} A_j\right)
=
A_i - \sum_{j=i+1}^{k} A_j.
\end{displaymath}

Now
\begin{equation}
\label{eq-Ai-recurrence}
A_i 
=
2^{k-i}(mk+1) 
= (mk+1) + \sum_{j=i+1}^{k} 2^{k-j}(mk+1) 
= (mk+1) + \sum_{j=i+1}^{k} A_j,
\end{equation}
so $A_i - \sum_{j=i+1}^k A_j = mk+1$.
But the ratio between the weight of $b$ and its corresponding good
path $a$ is then at most $2^{A(v^b)-A(v^a)} = 2^{-mk-1}$.  
Since there are only $2^{mk}$ paths altogether,
there are fewer than
$2^{mk}$ bad paths; thus,
the ratio between the total weight of all bad paths mapped to $a$ and
the weight of $a$ is less than $2^{mk} 2^{-mk-1} = \frac{1}{2}$.
Summing over all good paths shows that the total weight of all bad
paths is less than half the total weight of all good paths, so at
least $\frac{2}{3}$ of all non-abstaining paths are good.
It follows that, conditioned on not abstaining,
$M'$ accepts with probability greater than
$\frac{2}{3}$ if $M$ accepts, and accepts with probability less than
$\frac{1}{3}$ if $M$ rejects.  Hence $L(M) \in \bcpp$.
\end{proof}

An interesting open question is where exactly \bcpp{} lies between
\pnplog{} and \pp{}.  It is conceivable that by cleverly
exploiting the power of conditioning to amplify low-probability events
one could show $\bcpp = \pp$.
However, 
we will content ourselves with the much easier result of 
showing that the usual amplification
technique for \bpp{} also applies to \bcpp{}.

\begin{theorem}
\label{theorem-bcpp-amplification}
If $L \in \bcpp$, then there exists a noncommittal Turing machine $M$
such that the probability that $M$ accepts conditioned on not
abstaining is at least $1-f(n)$ if $x \in L$ and at most $f(n)$ if $x
\notin L$, where $n = |x|$ and $f(n)$ is any function of the form
$2^{-O(n^c)}$ for some constant $c > 0$.
\end{theorem}
\begin{proof}
Let membership in $L$ be computed by $M'$.
Assume $x \in L$ (the case $x \notin L$ is symmetric).
Consider $k$ independent executions of $M'$ with input $x$;
call the random variables representing their outputs $Y_1, Y_2,\ldots,Y_k$.
Because the executions are independent,
for any $0$-$1$ vector of values $y_i$
$\Pr[\forall i Y_i = y_i \mid \forall i Y_i \neq \perp]
 = \prod_i \Pr[Y_i = y_i \mid Y_i \neq \perp]$.
So conditioning on no abstentions,
$\sum_i Y_i$ has a binomial
distribution with $p = \frac{1}{2}+\epsilon$, and 
Chernoff bounds imply
$\Pr[\sum_i Y_i < \frac{k}{2} \mid \forall i Y_i \neq \perp]$
is exponentially small in $k\epsilon$.
Since $\epsilon$ is constant and we can make $k$ polynomially large
in $n$, the result follows.
\end{proof}

\subsubsection{Bounded Market Prediction Is \bcpp{}-complete}
\label{section-bcpp-complete}

In Section~\ref{section-bcpp}, we have defined the complexity class \bcpp{}
and have shown that it contains the powerful class $\pnplog$.  In this
section, we show that bounded market prediction is complete for
$\bcpp$.  In a sense, this result says that market prediction is a
universal prediction problem: if we can predict a market, we can
predict any event conditioned on past history as long as we can sample
from an underlying discrete probability space whose size is at most
exponential.

It also says that bounded market prediction is very hard. That is, using
Theorems~\ref{theorem-bcpp-amplification} and
\ref{theorem-pnplog-subset-bcpp}, even if
the next day's price is determined with all but an exponentially small
probability, it cannot be solved in the polynomial-time hierarchy
unless the hierarchy collapses to a finite level.

\begin{theorem}
\label{theorem-bcpp-complete}
The bounded market prediction problem is complete for \bcpp, in either
the AS+FI or the AS+PI model.
\end{theorem}
\begin{proof}
First we show that bounded market prediction is a member of
\bcpp{}.  Given a market, construct a noncommittal Turing machine $M$
whose input is the price history and strategies, and whose random
inputs supply the settings for the population variables $X_i$.
Let $M$ abstain if the price history is inconsistent with the input
and population variables;
depending on the model,
this is either a matter of checking the linear inequalities produced
by Lemma~\ref{lemma-FI-market-to-inequalities}
or the equations produced
by Lemma~\ref{lemma-PI-market-to-equations}.  
Otherwise, $M$ accepts
if the market rises and rejects if the market falls on the next day.
The probability that $M$ accepts thus equals the probability that
the market rises: either more than $2/3$ or less than $1/3$.
Since the problem is to distinguish between these two cases,
$M$ solves the problem within the definition of a $\bcpp$-machine.

In the other direction, 
we reduce from any \bcpp-language $L$.
Suppose $L$ is accepted by some \bcpp-machine $M$.
We will translate
$M$ and its input $x$ into a bounded market prediction
problem.  First use Theorem~\ref{theorem-bcpp-amplification} to
amplify the conditional probability that $M$ accepts to either more
than $2/3$ or less than $1/3$ as bounded market prediction demands.
Then convert $M$ into two polynomial-size circuits, one computing
\begin{displaymath}
C_{\not\perp}(r) = \left\{
\begin{array}{ll}
0 & \mbox{if $M(x,r) = \perp$}; \\
1 & \mbox{if $M(x,r) \neq \perp$},
\end{array}
\right.
\end{displaymath}
and the other computing
\begin{displaymath}
C_1(r) = \left\{
\begin{array}{ll}
0 & \mbox{if $M(x,r) \neq 1$}; \\
1 & \mbox{if $M(x,r) = 1$}.
\end{array}
\right.
\end{displaymath}

Without loss of generality we may assume that $C_{\not\perp}$ and $C_1$ 
are built from NOR gates.  Applying Lemma
\ref{lemma-circuit-to-inequalities} to each yields
two sets of constraints $A_{\not\perp}y > 0$ and $A_1 y > 0$
and column vectors $c_{\not\perp}$ and $c_1$
such that $c_{\not\perp}y > 0$ if and only if $C_{\not\perp}y = 1$
and $c_1 x > 0$ if and only if $C_1(x) = 1$, where $y$ satisfies the
previous linear constraints and $x$ is the initial prefix of $y$
consisting of variables not introduced by the construction of Lemma
\ref{lemma-circuit-to-inequalities}.
We also have from Lemma~\ref{lemma-circuit-to-inequalities}
that there is a one-to-one correspondence between assignments of $x$
and assignments of $y$ satisfying the $A$ constraints, so
probabilities are not affected by this transformation.

Now use Lemma~\ref{lemma-inequalities-to-FI-market} to construct a market
model in which $A_{\not\perp}y > 0$, $A_1 y > 0$, and $c_{\not\perp} y
> 0$ are enforced by the strategies and price history, and
$\sgn(c_1 y)$ determines the price change on the next day of trading.
Thus the consistent settings of the variables $X_i$ are precisely those
corresponding to settings of $r$ for which $C_{\not\perp}(r) = 1$, or,
in other words, those yielding computation paths that do not abstain.
The market rises when $C_1(r) = 1$, or when $M$ accepts.  So if we can
predict whether the market rises or falls with conditional probability
at least $2/3$, we can predict the likely output of $M$.  It follows
that bounded market prediction for the AS+FI model is \bcpp-hard.

To show the similar result for the AS+PI model, use
Lemma~\ref{lemma-slack-variables} to convert the constraints
$A_{\not\perp}y > 0$, $A_1 y > 0$ into a system of linear equations
$Bz = 1$, and then proceed as before, using
Lemma~\ref{lemma-equations-to-PI-market} to convert this system to a
price history and letting $c_1 z$ determine the price change (and thus
the sign of the price change) on the next day of trading.
\end{proof}

\subsubsection{Unbounded Market Prediction is \cpp{}-complete}
\label{sec_unbounded}
The unbounded market prediction problem seems harder because the
probability threshold in question is $\frac{1}{2}$ with no $\epsilon$
bound in contrast to the thresholds $\frac{2}{3}$ and $\frac{1}{3}$
for the bounded market prediction problem. The following theorem
reflects this intuition.  However, since we do not know whether
\bcpp{} is distinct from \pp{}, we do not know whether unbounded
prediction is strictly harder.

\begin{theorem}\label{theorem-cpp-complete}
The unbounded market prediction problem is complete for $\cpp = \pp$,
in either the AS+FI or the AS+PI model.
\end{theorem}
\begin{proof}
Similar to the proof of Theorem~\ref{theorem-bcpp-complete}.
\end{proof}

\section{Future Research Directions}\label{sec_conclusion}

There are many problems left open in this paper. Below we briefly
discuss some general directions for further research.

We have reported a number of simulation and theoretical results for
the AS model.  As for empirical analysis, it would be of interest to
fit actual market data to the model. We can then use the estimated
parameters to (1) test whether the model has any predicative power and
(2) test the effectiveness of new or known trading algorithms.  This
direction may require carefully choosing ``{realistic}''
strategies for $\Pi$.  Besides the momentum and contrarian strategies,
there are some popular ones which are worth considering, such as those
based on support levels.  Investment newsletters could be a useful
source of such strategies.

The AS model is an idealized one.  We have chosen such simplicity as a
matter of research methodology.  It is relatively easy to design
highly complicated models which can generate very complex market
behavior. A more challenging and interesting task is to design the
simplest possible model which can generate the desired market
characteristics.  For instance, a significant research direction would
be to find the simplest model in which market prediction is
computationally hard.  On the other hand, it would be of great
interest to find the most general models in which market prediction
takes only polynomial time.  For this goal, we can consider injecting
more realism into the model by introducing resource-bounded learning
(the generality of $\Pi$ is equivalent to unbounded learning),
variable memory size, transaction costs, buying
power, limit orders, short sell, options, etc.

\section*{Acknowledgments}
This work originated with David Fischer's senior project in 1999,
advised by Ming-Yang Kao.  David would like to thank his father and
role model, Professor Michael Fischer, for teaching, mentoring, and
inspiring him throughout college.

\bibliographystyle{abbrv}
\bibliography{prediction} 

\end{document}